\documentclass[twocolumn,showpacs,preprintnumbers,amsmath,amssymb,superscriptaddress]{revtex4-1}

\usepackage{graphicx}
\usepackage{dcolumn}
\usepackage{bm}

\begin{document}


\title{A structural distortion induced magneto-elastic locking in Sr$_2$IrO$_4$ revealed through nonlinear optical harmonic generation}

\author{D. H. Torchinsky}
\affiliation{Institute for Quantum Information and Matter, California Institute of Technology, Pasadena, CA 91125, USA}
\affiliation{Department of Physics, California Institute of Technology, Pasadena, CA 91125, USA}
\author{H. Chu}
\affiliation{Institute for Quantum Information and Matter, California Institute of Technology, Pasadena, CA 91125, USA}
\affiliation{Department of Applied Physics, California Institute of Technology, Pasadena, CA 91125, USA}
\author{L. Zhao}
\affiliation{Institute for Quantum Information and Matter, California Institute of Technology, Pasadena, CA 91125, USA}
\affiliation{Department of Physics, California Institute of Technology, Pasadena, CA 91125, USA}
\author{N. B. Perkins}
\affiliation{Department of Physics, University of Wisconsin, Madison, WI 53706, USA}
\author{Y. Sizyuk}
\affiliation{Department of Physics, University of Wisconsin, Madison, WI 53706, USA}
\author{T. Qi}
\affiliation{Center for Advanced Materials, Department of Physics and Astronomy, University of Kentucky, Lexington, Kentucky 40506, USA}
\author{G. Cao}
\affiliation{Center for Advanced Materials, Department of Physics and Astronomy, University of Kentucky, Lexington, Kentucky 40506, USA}
\author{D. Hsieh}
\affiliation{Institute for Quantum Information and Matter, California Institute of Technology, Pasadena, CA 91125, USA}
\affiliation{Department of Physics, California Institute of Technology, Pasadena, CA 91125, USA}

\date{\today}

\begin{abstract}
We report a global structural distortion in Sr$_2$IrO$_4$ using spatially resolved optical second and third harmonic generation rotational anisotropy measurements. A symmetry lowering from an $I4_{1}/acd$ to $I4_{1}/a$ space group is observed both above and below the N\'{e}el temperature that arises from a staggered tetragonal distortion of the oxygen octahedra. By studying an effective super-exchange Hamiltonian that accounts for this lowered symmetry, we find that perfect locking between the octahedral rotation and magnetic moment canting angles can persist even in the presence of large non-cubic local distortions. Our results explain the origin of the forbidden Bragg peaks recently observed in neutron diffraction experiments and reconcile the observations of strong tetragonal distortion and perfect magneto-elastic locking in Sr$_2$IrO$_4$.
\end{abstract}

\maketitle

Iridium oxides are predicted to realize a variety of exotic quantum phases \cite{Krempa_Review,Jackeli,Lawler,Machida,Jiang,Wan,Pesin,YangKim,Maciejko} that emerge from a rare combination of strong electron-electron repulsion, spin-orbit coupling (SOC) and crystalline electric field (CEF) splitting. Over the past several years, one of the most intensively studied iridates has been Sr$_2$IrO$_4$ owing to its novel $J_{eff}=1/2$ Mott insulating ground state \cite{Kim_PRL,Kim_Science} and the similarity of its crystallographic, electronic and magnetic structures to the parent compound La$_2$CuO$_4$ of the high-$T_c$ cuprates. However, despite predictions of unconventional superconductivity in Sr$_2$IrO$_4$ upon chemical doping \cite{Fa,Martins,Watanabe_PRL,Meng} and the observation of Fermi arcs with a pseudogap behavior \cite{Kim_arc}, it is unclear why no signatures of superconductivity are detected. Very recent experiments suggest that the structural and magnetic properties of Sr$_2$IrO$_4$ are in fact not completely understood. In particular, neutron diffraction studies on Sr$_2$IrO$_4$ single crystals report new Bragg peaks \cite{Ye,Dhital} that challenge its long accepted crystal structure \cite{Crawford,Huang,Kim_Science} while resonant x-ray diffraction studies report a near perfect locking of the magnetic moment canting and oxygen octahedra rotation angles \cite{Kim_Science,Boseggia_JPCM} that cannot be fully explained by existing theoretical models \cite{Jackeli,perkins14}. A more detailed understanding of the structural and magnetic properties of Sr$_2$IrO$_4$ is therefore imperative to resolving these outstanding questions.

Iridates, in general, pose certain challenges for diffraction based structure determination probes. Available single crystals are typically small ($\leq1~$mm) and exhibit domains \cite{Ye,Dhital,Boseggia327} that require highly localized ($\leq100~\mu$m) probe beams to isolate \cite{Boseggia327}. Subtle distortions of the oxygen lattices, which are especially prevalent among the Ruddlesden-Popper series Sr$_{n+1}$Ir$_n$O$_{3n+1}$, are difficult to resolve due to the weak x-ray scattering cross section of oxygen \cite{Kim_Science,Clancy,Boseggia327}. Neutron scattering signals are also weak due to the strong absorption cross section of iridium and its small magnetic moment.

\begin{figure}
\includegraphics[scale=0.42,clip=true, viewport=0.0in 0.0in 8.8in 9.4in]{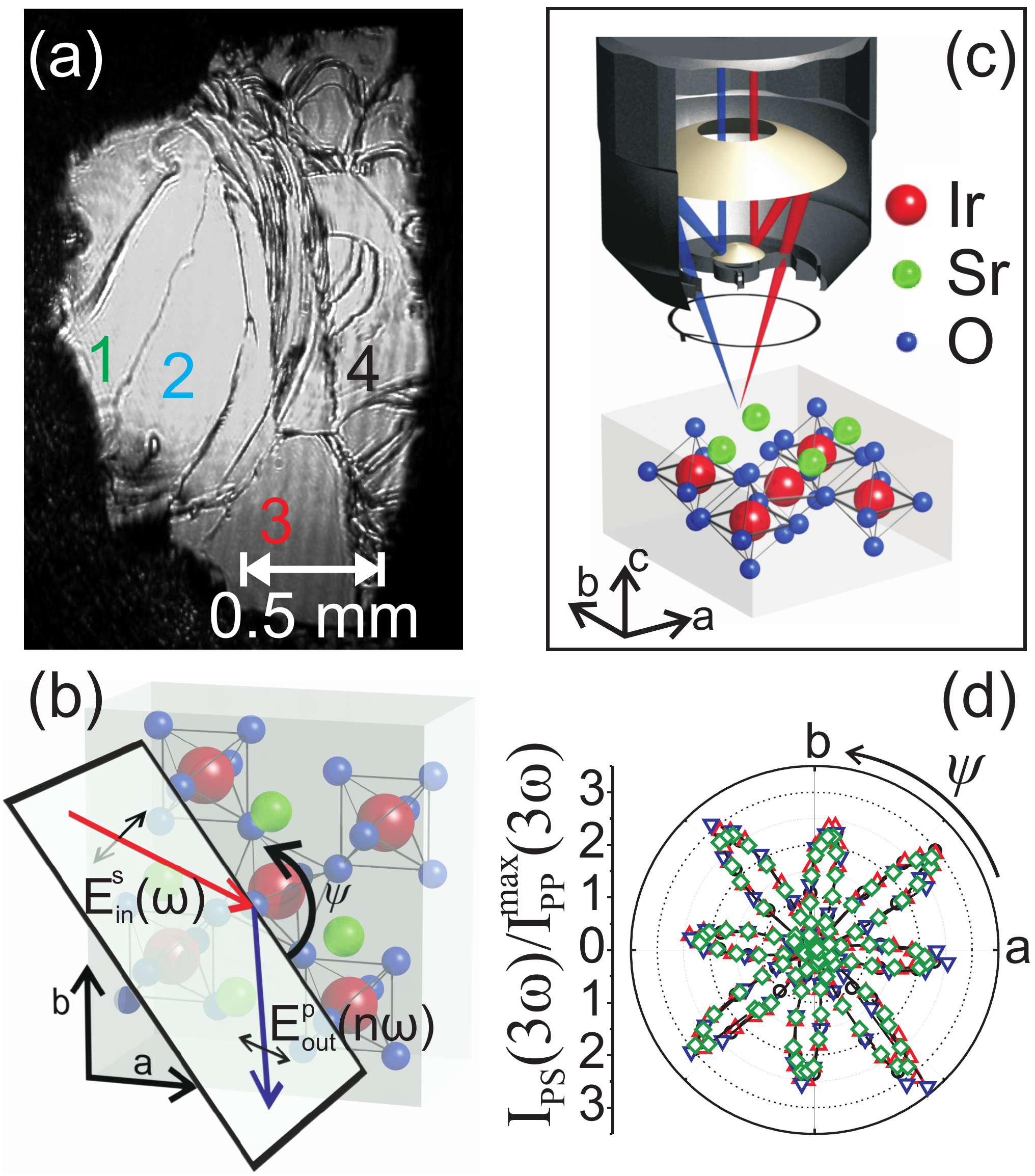}
\caption{\label{fig:Fig1} (Color online) (a) Typical optical image of the (001) surface of a Sr$_2$IrO$_4$ single crystal. (b) Schematic of the RA experimental geometry. P(S)-polarization denotes an electric field lying parallel (perpendicular) to the scattering plane. (c) Layout of the scanning RA setup showing how the rotating scattering plane is created inside the reflective objective lens \cite{Torchinsky_RSI}. (d) Select THG-RA patterns taken in PS geometry obtained from regions 1-4 in panel (a).}
\end{figure}

In this Letter, we report a global bulk structural distortion in Sr$_2$IrO$_4$ observed using a combination of spatially resolved optical second harmonic generation (SHG) and third harmonic generation (THG) experiments. Our technique is highly sensitive to small changes in bulk symmetry and is able to probe micron sized areas of a crystal (Fig.~\ref{fig:Fig1}a), thus providing complementary information to neutron and x-ray diffraction. By studying an effective super-exchange Hamiltonian, we show that these newfound broken symmetries introduce modifications to the $J_{eff}=1/2$ model that naturally explain the robust locking of the moment canting and oxygen octahedra rotation angles.

Nonlinear optical harmonic generation is a process by which light of frequency $\omega$ is converted into higher harmonics $n\omega$ ($n$ = 2,3,4...) through its nonlinear interaction with a material \cite{Shen}. By Neumann's principle, the nonlinear optical susceptibility tensors that relate the incident electric field $\vec{E}$ to induced electric dipole $P_i(n\omega)=\chi^{ED}_{ijk...}E_j(\omega)E_k(\omega)...$, electric quadrupole $Q_{ij}(n\omega)=\chi^{EQ}_{ijkl...}E_k(\omega)E_l(\omega)...$, magnetic dipole $M_i(n\omega)=\chi^{MD}_{ijk...}E_j(\omega)E_k(\omega)...$ or even higher order multipole densities, which act as sources of higher harmonic radiation, must be invariant under every symmetry operation of the crystal~\cite{Birss}. These conditions of invariance establish a set of relationships between tensor components that reduce the number of independent non-zero components. The structure of $\tensor{\chi}$ therefore encodes the symmetries of a crystal, with higher rank $\tensor{\chi}$ allowing for more accurate levels of refinement.


The components of $\tensor{\chi}$ can be measured through rotational anisotropy (RA) experiments where the intensity of high harmonic light reflected from a crystal is recorded as a function of the angle $\psi$ subtended between the light scattering plane and a crystalline axis (Fig.~\ref{fig:Fig1}b). Since RA patterns measured using different combinations of incident and reflected light polarization are sensitive to different sets of tensor components, a collection of RA patterns is typically required to completely determine the structure of $\tensor{\chi}$. Unlike conventional RA setups that mechanically rotate the sample \cite{Tom1,Heinz,Nyvlt,Hsieh_SHG}, we have developed a rotating scattering plane based approach that allows micron sized regions to be scanned over in a vacuum cryostat \cite{Torchinsky_RSI}. For our experiments, an optical parametric amplifier pumped by a regenerative Ti:sapphire amplifier produces wavelength tunable laser pulses with $<$100~fs duration at a 10~kHz repetition rate. The beam is focused down to a $\sim20~\mu$m spot on the sample surface at an oblique incidence angle of $\sim30^{\circ}$ using a reflective objective (Fig.~\ref{fig:Fig1}c). Less than 1 mW average incident power was used in order to avoid photoinduced sample damage. Crystals were oriented using x-ray Laue diffraction. Details about the growth and characterization of Sr$_2$IrO$_4$ single crystals are published elsewhere \cite{Cao}.

Previous works have assigned Sr$_2$IrO$_4$ to a centrosymmetric tetragonal $4/mmm$ crystallographic point group ($I4_{1}/acd$ space group) \cite{Crawford,Huang,Kim_Science}. The reported structure is composed from layers of corner sharing IrO$_6$ octahedra, which exhibit a uniform tetragonal distortion arising from an elongation of the octahedra along the $c$-axis and a staggered rotation that creates a two-sublattice structure. However recent single crystal neutron diffraction studies observe additional nuclear Bragg peaks that violate the $I4_{1}/acd$ space group \cite{Ye,Dhital}. These forbidden peaks may originate from structural defects such as oxygen vacancies that distort the local symmetry or from a subtle global symmetry reduction. Although a number of alternative centrosymmetric and non-centrosymmetric space groups have been proposed, diffraction experiments currently cannot distinguish between them \cite{Ye,Dhital}.

To examine the possibility of local symmetry variations, we performed scanning THG-RA measurements with $\sim20 \mu$m spatial resolution on (001) cleaved surfaces of Sr$_2$IrO$_4$ (Fig.~\ref{fig:Fig1}a), which is sensitive to the local bulk crystal symmetry as we will discuss later. We observed no changes in the magnitude or symmetry of the THG-RA patterns across the entire surfaces of several crystals (Fig.~\ref{fig:Fig1}d), which suggests that the entire crystal likely belongs to a lower symmetry subgroup of $4/mmm$. In order to determine the subgroup we first performed SHG-RA measurements, which are particularly well suited to distinguishing between centrosymmetric and non-centrosymmetric point groups because the usually dominant $\chi^{ED}_{ijk}$ (odd rank) contribution to SHG vanishes under inversion symmetry \cite{Shen,Hirata}.

\begin{figure}
\includegraphics[scale=0.66,clip=true, viewport=0.0in 0in 8.8in 10.4in]{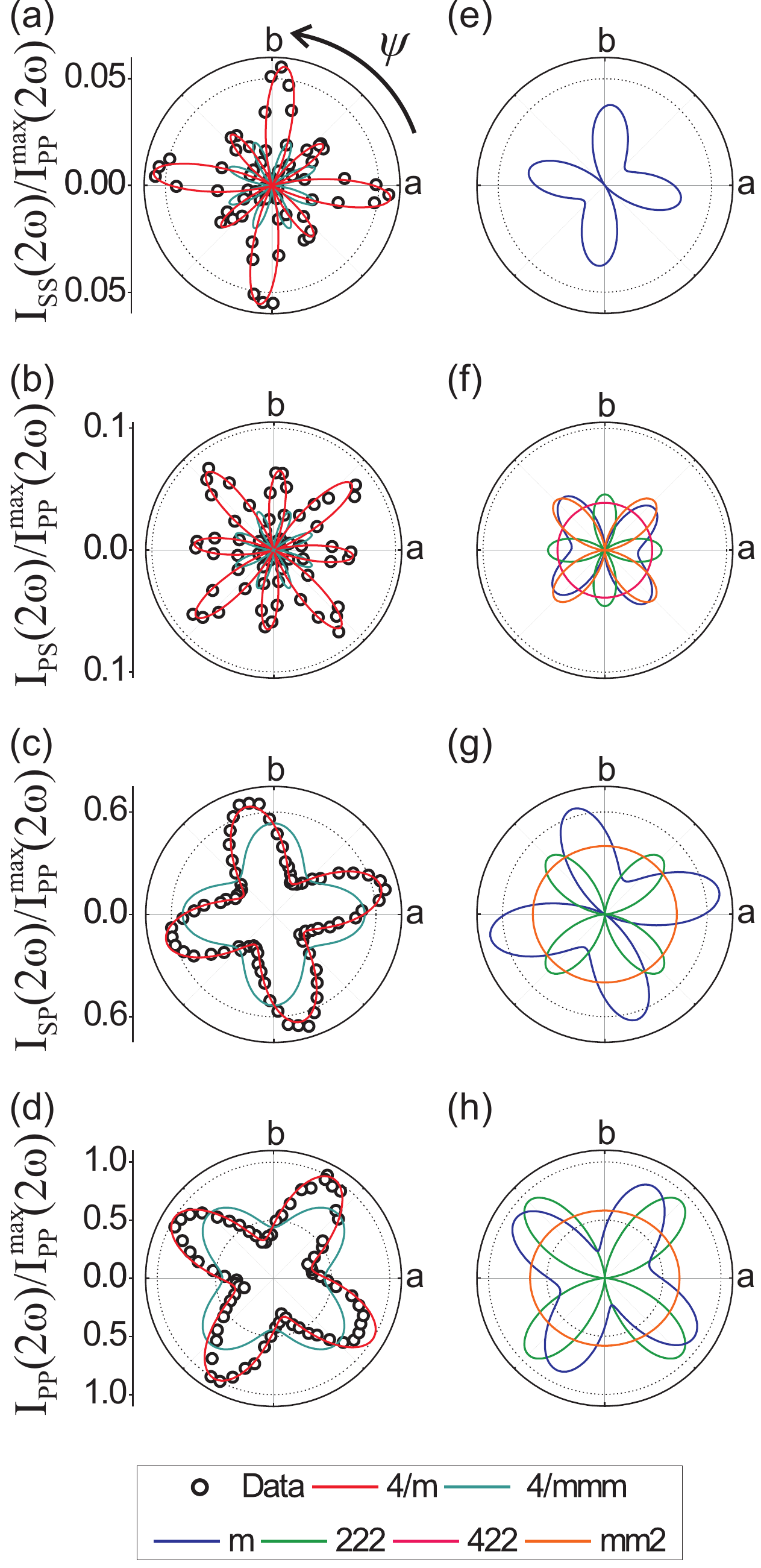}
\caption{\label{fig:Fig2} (Color online) SHG-RA patterns (open circles) taken under (a) SS, (b) PS, (c) SP and (d) PP geometries. The data were collected using 800~nm incident and 400~nm reflected light at $T=295$~K. The magnitudes of all patterns are normalized to the PP trace. Red and cyan lines are best fits to bulk electric quadrupole induced SHG-RA calculated using centrosymmetric $4/m$ and $4/mmm$ point groups respectively. (e)-(h) show best fit results to bulk electric dipole induced SHG calculated using the three proposed non-centrosymmetric point groups $mm2$, $222$ and $422$ as well as the monoclinic point group $m$ for comparison. Responses that are absent in the plots are symmetry forbidden.}
\end{figure}

Figures~\ref{fig:Fig2}(a) to (d) show SHG-RA patterns collected under all four distinct linear polarization combinations and figures~\ref{fig:Fig2}(e) to (h) show best fits to bulk electric dipole induced SHG calculated using the three non-centrosymmetric subgroups that have been proposed in the literature \cite{Ye,Dhital}: orthorhombic $mm2$ (space group $Pnn2$), orthorhombic $222$ (space group $I2_{1}2_{1}2_{1}$) and tetragonal $422$ (space group $I4_1$22). Results using the non-centrosymmetric monoclinic subgroup $m$ are also plotted for comparison. It is clear that the data cannot be described by any of these non-centrosymmetric subgroups. A centrosymmetric tetragonal subgroup $4/m$ (space group $I4_{1}/a$) has also been proposed \cite{Ye}, however like the case for $4/mmm$, bulk electric dipole induced SHG is forbidden. Rather a bulk electric quadrupole induced SHG process must be responsible \cite{EPAPS} in these cases and fits to both $4/m$ and $4/mmm$ are overlayed in figures~\ref{fig:Fig2}(a)-(d). Overall the data are clearly better described by a $4/m$ rather than a $4/mmm$ point group or any of the other non-centrosymmetric subgroups. Most importantly, the rotation of the peaks and valleys of all patterns away from the high symmetry directions of the crystal indicates an absence of mirror symmetry about the $ac$, $bc$ or the diagonal planes, which is consistent with a $4/m$ but not with a $4/mmm$ point group. Furthermore, the mathematical expressions for the bulk electric quadrupole induced PS and SS patterns derived using a $4/mmm$ point group, which are both proportional to $|\sin(4\psi)|^2$, yield an eight-fold rotational symmetric pattern that cannot explain the clear modulations observed in the lobe amplitudes (Fig.~\ref{fig:Fig2}a \& b). On the other hand the corresponding expressions derived using a $4/m$ point group, which are both proportional to $|\eta_1+\eta_2 \cos(4\psi)+\eta_3\sin(4\psi)|^2$ where $\eta_{1,2,3}$ are linear combinations of $\chi^{EQ}_{ijkl}$ tensor components \cite{EPAPS}, do allow for such modulations.

\begin{figure}
\includegraphics[scale=0.58,clip=true, viewport=-0.1in 0in 8.6in 5.5in]{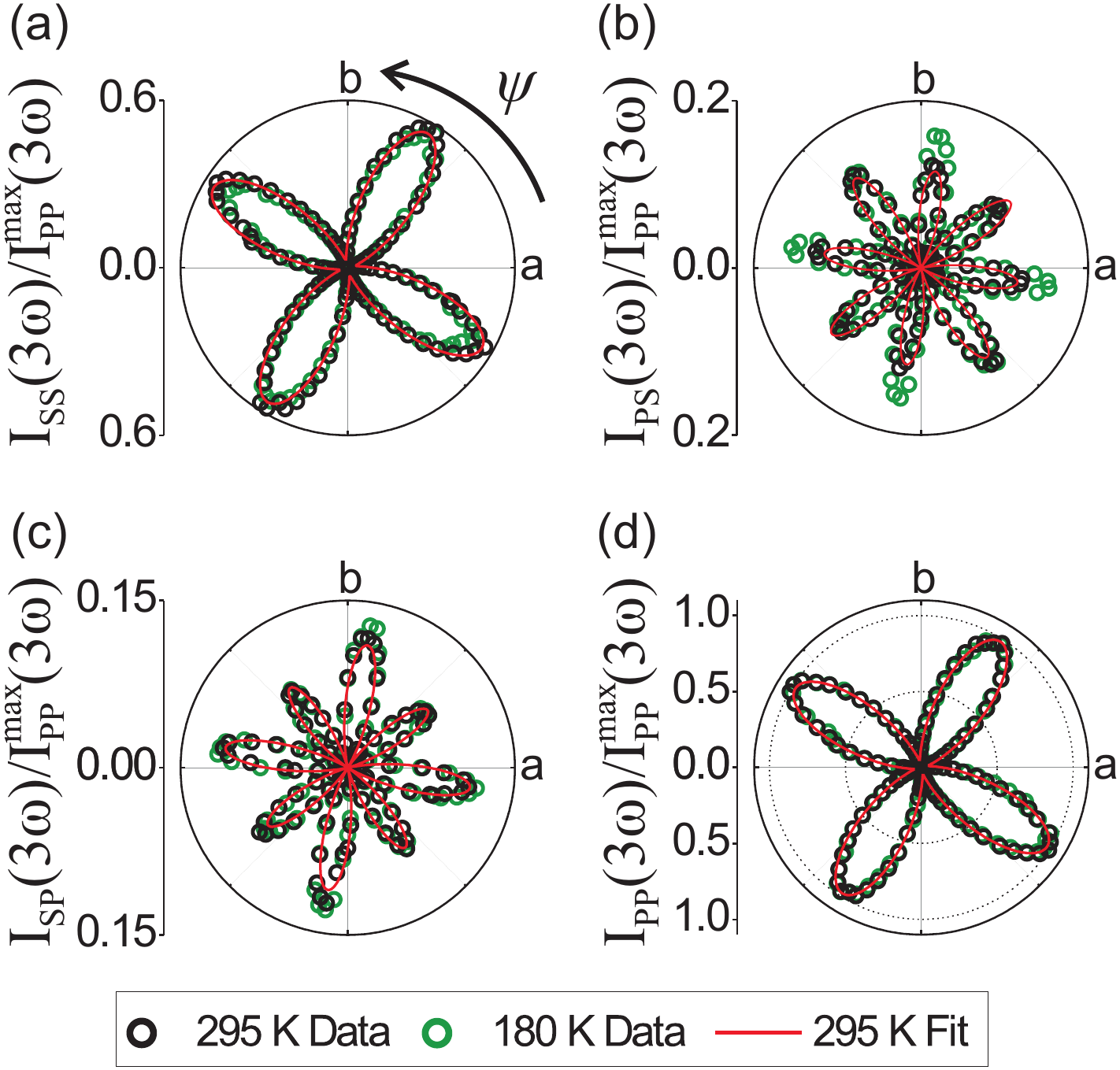}
\caption{\label{fig:Fig3} (Color online) THG-RA patterns taken under (a) SS, (b) PS, (c) SP and (d) PP geometries. The data were collected using 1200~nm incident and 400~nm reflected light at both $T=295$~K (black circles) and $T=180$~K (green circles). The magnitudes of all patterns are normalized to the PP trace. Red lines are best fits to the 295~K data using bulk electric dipole induced THG-RA calculated using centrosymmetric $4/m$ point group.}
\end{figure}

Although the SHG-RA data are most consistent with a $4/m$ point group out of all point groups proposed by diffraction based studies, we cannot completely rule out the possibility that the SHG-RA patterns arise from a coherent sum of bulk electric quadrupole and bulk magnetic dipole or surface electric dipole contributions to SHG, which can be comparable in magnitude \cite{Shen}. On the other hand, THG-RA measurements are established bulk sensitive probes of centrosymmetric crystals \cite{Sipe,Lafrentz} because the bulk electric dipole contribution $\chi^{ED}_{ijkl}$ (even rank) is allowed. Figure~\ref{fig:Fig3} shows THG-RA patterns with best fits to bulk electric dipole induced THG from a $4/m$ point group overlayed. Similar to the SHG data (Fig.~\ref{fig:Fig2}), the peaks and valleys of the THG-RA patterns are rotated away from the high symmetry directions of the crystal, which cannot be reproduced by fits to a $4/mmm$ point group \cite{EPAPS}. Moreover, there is a modulation of the lobe amplitudes in the PS and SP geometry data that can only be described using a $4/m$ point group. The mathematical expression for the PS and SP patterns derived using a $4/mmm$ point group, which is proportional to $|\sin(4\psi)|^2$, cannot account for these features \cite{EPAPS}.

\begin{figure}
\includegraphics[scale=0.395,clip=true, viewport=0.0in 0in 8.8in 8.5in]{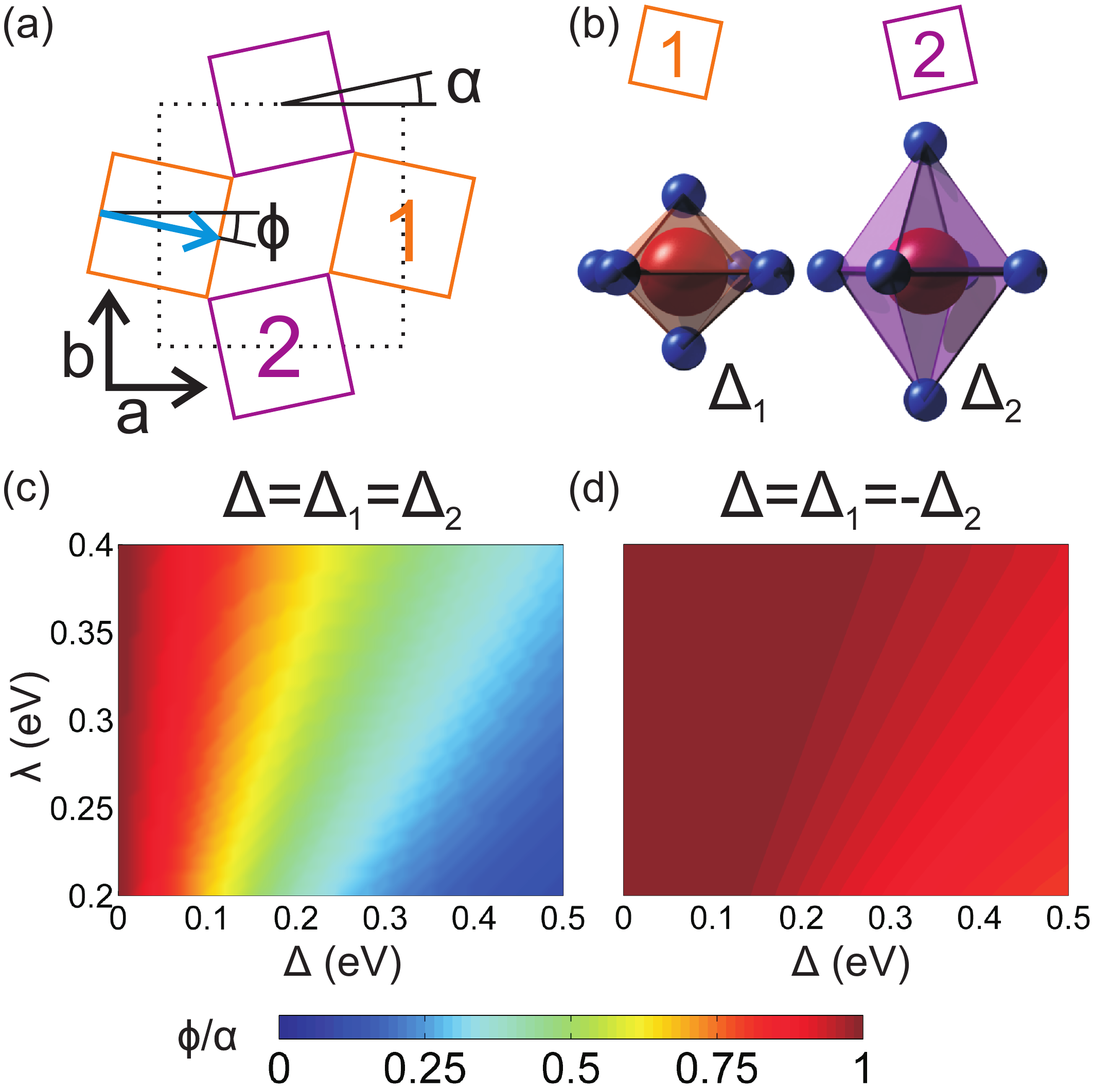}
\caption{\label{fig:Fig4} (Color online) (a) Illustration of an IrO$_2$ plane in Sr$_2$IrO$_4$. The oxygen octahedra rotate about the $c$-axis by $\pm\alpha$ creating a two sublattice structure. The magnetic moments couple to the lattice and exhibit canting angles $\pm\phi$. (b) An unequal tetragonal distortion ($\Delta_1$ and $\Delta_2$) on the two sublattices as required by the $I4_{1}/a$ space group. (c) The ratio $\phi/\alpha$ as a function of both $\lambda$ and $\Delta$ calculated for the case of uniform and (d) staggered ($\Delta_1=-\Delta_2$) tetragonal distortion using eqn.(\ref{hamtet}) assuming a Coulomb energy $U_1$=2.4 eV, a Hund's coupling $J_H$=0.3 eV, hopping $t$=0.13 eV and $\alpha$=11.5$^{\circ}$.}
\end{figure}

The observed rotation of the peaks and valleys away from the high symmetry directions of the crystal and modulation of lobe amplitudes in both the SHG-RA and THG-RA patterns together show that Sr$_2$IrO$_4$ exhibits a globally reduced bulk structural symmetry that is best described by the $I4_{1}/a$ space group. This implies that the $c$- and $d$-glide planes previously thought to exist in the $I4_{1}/acd$ description are actually absent, which can only occur if the tetragonal distortions of the oxygen octahedra on the two sublattices are inequivalent (Fig.~\ref{fig:Fig4}a \& b) \cite{EPAPS}. Owing to the strong magneto-elastic coupling in Sr$_2$IrO$_4$ \cite{Ge,Chikara,Haskel}, this staggered tetragonal distortion, which is present in our data both above and below the N\'{e}el temperature $T_{N}=240$~K (Fig.~\ref{fig:Fig3}), will likely influence how the magnetic moments couple to the octahedral rotations.

The most widely used model \cite{Jackeli} for understanding the relationship between the moment canting angle $\phi$ and the octahedral rotation angle $\alpha$ (Fig.~\ref{fig:Fig4}a) was developed by Jackeli and Khaliullin (JK) assuming an $I4_{1}/acd$ space group, which only allows for a uniform tetragonal distortion. In the JK model, the ratio $\phi/\alpha$ depends only on a parameter $\theta$ defined by $\tan(2\theta)=2\sqrt{2}\lambda/(\lambda-2\Delta)$, where $\lambda$ and $\Delta$ are the strengths of SOC and uniform tetragonal CEF splitting, respectively. A perfect magneto-elastic locking ($\phi/\alpha=1$) is predicted in the cubic limit ($\Delta$=0) where the local magnetic degrees of freedom are derived from a $J_{eff}=1/2$ Kramers doublet. Any mixing between the $J_{eff}=1/2$ and $J_{eff}=3/2$ states introduced through tetragonal distortion causes $\phi/\alpha$ to be either smaller or larger than 1 depending on whether the oxygen octahedra are elongated ($\Delta>0$) or compressed ($\Delta<0$). Using commonly accepted \cite{Kim_PRL,Jackeli,Jin,perkins14,Zhang} values of $\lambda$ ($\sim400$ meV) and $\Delta$ ($\sim140$ meV) or their experimentally derived ratio ($\Delta/\lambda\sim0.34$) \cite{Haskel} in Sr$_2$IrO$_4$, the JK model predicts that $\phi/\alpha \approx 0.7$. This, however, is incompatible with recent neutron and resonant x-ray diffraction studies that report values of $\phi$ = 13(1)$^{\circ}$ \cite{Ye} and 12.2(8)$^{\circ}$ \cite{Boseggia_JPCM} and $\alpha$ = 11.8(1)$^{\circ}$ \cite{Ye}, which indicate a nearly perfect magneto-elastic locking.

To investigate how perfect magneto-elastic locking can remain robust even under substantial departure from cubic symmetry, we developed an extension of the JK model that accounts for a staggered tetragonal distortion. The starting point of our model is a microscopic single-ion Hamiltonian that includes both SOC and tetragonal CEF distortion ~\cite{perkins14}. Its ground state is a Kramers doublet whose orbital and spin composition is determined by the relative strengths of SOC and tetragonal splitting, which will depart from a $J_{eff}=1/2$ description for any non-zero value of the tetragonal splitting. By allowing for unequal tetragonal splitting ($\Delta_1$ and $\Delta_2$) on the two sublattices, the doublets on each sublattice will in general possess different spin and orbital compositions. We treat the resulting doublets as pseudospin-1/2 degrees of freedom ($\bf{S}$) that interact via the following super-exchange Hamiltonian derived in Ref.~\cite{perkins14}:
\begin{eqnarray}
\label{hamtet}
&&H= \Sigma_{n,n'} J {\bf S}_n {\bf S}_{n^{\prime}}-D(S_{n}^{x}S_{n^{\prime}}^{y}-S_{n}^{y}S_{n^{\prime}}^{x})
\\\nonumber
&&+\delta J_z S^z_n  S^z_{n^{\prime}}
+\delta J_{xy}\left({\bf S}_n \cdot {\bf r}_{n,n^{\prime}}\right)
 \left({\bf S}_{n^{\prime}} \cdot{\bf r}_{n,n^{\prime}}\right)~
\end{eqnarray}\\
where the isotropic exchange $J$, exchange anisotropies $\delta J_z$ and $\delta J_{xy}$, and Dzyaloshinsky-Moriya interaction $D$ are functions of the microscopic parameters $\lambda$, $\Delta_1$, $\Delta_2$, Coulomb interaction and Hund's coupling. The $x$ and $y$ axes point along the $a$ and $b$ directions, respectively, and ${\bf r}_{n,n^{\prime}}$ is the unit vector along the $n,n^{\prime}$ bond. Classical minimization of the super-exchange Hamiltonian is employed to calculate the moment canting angles. These methods are fully described in Ref.\cite{perkins14}.

In the case of a uniform tetragonal distortion ($\Delta_1$=$\Delta_2\equiv\Delta$) as required for an $I4_{1}/acd$ space group, we find that for a fixed value of $\lambda$, $\phi/\alpha$ sharply decreases from 1 as a function of $\Delta$ consistent with the JK model (Fig.~\ref{fig:Fig4}c). However if the sign of the tetragonal distortion is staggered between sublattices ($\Delta_1$=$-\Delta_2\equiv\Delta$), which is consistent with an $I4_{1}/a$ space group, then $\phi/\alpha$ becomes remarkably insensitive to both $\lambda$ and $\Delta$ (Fig.~\ref{fig:Fig4}d). This shows that the magnitude of $\phi/\alpha$ is more strongly influenced by the spatially averaged value of the tetragonal distortion rather than its local value on an individual oxygen octahedron, which allows the existence of a large local tetragonal distortion to be reconciled with the observation of perfect magneto-elastic locking.

In conclusion, our SHG-RA and THG-RA measurements reveal a globally lowered bulk structural symmetry in Sr$_2$IrO$_4$ that is induced by a staggered tetragonal distortion of the oxygen octahedra. Although we currently cannot obtain a quantitative measure of $\Delta_1$ and $\Delta_2$, we propose that a staggering of the sign of tetragonal CEF splitting naturally explains the observations of perfect magneto-elastic locking in the presence of non-cubic structural distortions. Quantitative measures of $\Delta_1$ and $\Delta_2$ using other techniques will be important for understanding the detailed spin and orbital composition of the ground state doublet \cite{Chapon,Sala} and the robustness of a $J_{eff}=1/2$ description to these lattice distortions \cite{Boseggia_PRL}. More generally, we have demonstrated a technique to perform symmetry refinement on micron length scales that can be highly complementary to diffraction based probes especially for the study of 5$d$ transition metal oxides.

\begin{acknowledgments}
We acknowledge useful discussions with Feng Ye, Bryan Chakoumakos, Stephen Lovesey, Dmitry Khalyavin, Peter Woelfle and Stephen Wilson. D.H. acknowledges partial support by the U. S. Army Research Office under grant number W911NF-13-1-0059. Instrumentation for the NHG-RA setup was partially supported by a U. S. Army Research Office DURIP award under grant number W911NF-13-1-0293. D.H. acknowledges funding provided by the Institute for Quantum Information and Matter, an NSF Physics Frontiers Center (PHY-1125565) with support of the Gordon and Betty Moore Foundation through Grant GBMF1250. N.P. and Y.S. acknowledge support from NSF grant DMR-1255544. G.C. acknowledges NSF support via grant DMR-1265162.
\end{acknowledgments}


\begin{thebibliography}{h}
\bibitem{Cao}
Single crystals of Sr$_2$IrO$_4$ were grown using a self-flux technique from off-stoichiometric quantities of IrO$_2$, SrCO$_3$ and SrCl$_2$. The ground mixtures of powders were melted at 1470$^{\circ}$C in partially capped platinum crucibles. The soaking phase of the synthesis lasted for $>$20 hours and was followed by a slow cooling at 2$^{\circ}$C/hr to reach 1400$^{\circ}$C. From this point the crucible is brought to room temperature through a rapid cooling at a rate of ~100$^{\circ}$C/hr.
\end{thebibliography}
\end{document}